\documentclass{mem}
\usepackage{txfonts}\usepackage{balance}
\usepackage{graphicx}
\usepackage[a4paper]{hyperref}
\idline{0}{0}
\begin{document}
\def\teff{$T\rm_{eff }$}
\def\kms{$\mathrm {km s}^{-1}$}

\title{Cosmographic applications of Gamma Ray Bursts}

   \subtitle{}

\author{
L. \,Izzo\inst{1,2}, O. \, Luongo\inst{1,2} \and S. \,
Capozziello\inst{1}   }

  \offprints{l.izzo@icra.it}

\institute{Dipartimento di Fisica, Universit\`a di Roma ''Sapienza''
and Icra, Piazzale Aldo Moro 5, I-00185 Roma, Italy.\and
Dipartimento di Scienze Fisiche, Universit\`a di Napoli ''Federico
II'' and INFN Sez. di Napoli, Compl. Univ. Monte S. Angelo, Ed. N,
Via Cinthia, I-80126 Napoli, Italy.\\ \email{luca.izzo@icra.it} }

\authorrunning{Izzo et al.}

\titlerunning{Cosmographic application of GRBs}

\abstract{ In this work we present some applications about the use
of the so-called Cosmography with GRBs. In particular, we try to
calibrate the Amati relation by using the luminosity distance
obtained from the cosmographic analysis. Thus, we analyze the
possibility of use GRBs as possible estimators for the cosmological
parameters, obtaining as preliminary results a good estimate of the
cosmological density parameters, just by using a GRB data sample.
}

\maketitle{}

\section{Introduction}

It is a matter of fact that Gamma Ray Bursts (GRBs) are the most
powerful explosions in the Universe; this feature makes them as one
of the most studied objects in high energy astrophysics. The flux
observed from their emission and the measurement of the redshift $z$
from the observations of the afterglow \cite{Frontera}, point out a
very high value for the isotropic energy emitted in the burst, so
that there are some GRBs observed at very high redshift. Up to date,
the farthest GRB has a spectroscopic redshift of $\sim$ 8.2
\cite{Tanvir,Salvaterra}. These interesting features suggest a
possible use of GRBs as distance indicators; unfortunately our
knowledge on the mechanisms underlying the GRB emission is not
completely understood, so that their use as standard candles is
still not clear. However, there exist some correlations among the
observed spectroscopic and photometric properties of the GRBs,
allowing us to put severe constraints on the GRBs distances. What we
need is an independent estimation of the isotropic energy $E_{iso}$
emitted from a GRB. Indeed, by using the GRB's fluence $S_{obs}$
measured by a detector in a certain energy range, it becomes
possible to determine the luminosity distance $d_l$ as follows
\begin{equation}\label{eq:no1}
d_l = \left(\frac{E_{iso}(1+z)}{S_{bolo}}\right)^{\frac{1}{2}},
\end{equation}
where $S_{bolo}$ is the bolometric fluence emitted, obtained from
the Schaefer formula \cite{Schaefer}
\begin{equation}\label{eq:no2}
S_{bol} = S_{obs} \frac{\int_{1/(1+z)}^{10^4/(1+z)} E \phi d
E}{\int_{E_{min}}^{E^{max}} E \phi d E}.
\end{equation}
In literature there are many correlation formulas\footnote{For a
review see \cite{Meszaros}.}, each of them takes in account
different observed quantities, but in this work we assume the
validity of the so-called Amati relation \cite{Amati}, for
different reasons
\begin{description}
  \item[1] it relates just the isotropic energy with the peak energy in the
$\nu$F($\nu$) spectrum without considering others,
  \item[2] it involves time-independent quantities, overcoming the problem of
  the large instrumental biases,
  \item[3] all of the long GRBs satisfy the Amati relation, while
the same is not true for the other correlations.
\end{description}
In addition, although the Amati relation suffers of some biases, as
the detector dependence of the observed quantity considered
\cite{Nemiroff,Butler}, it seems to be well verified from
observations \cite{Amati2}. However, one of the most relevant
challenge is represented by the calibration of the Amati relation,
because a low redshift sample of GRBs is, up to now, lacking; a
similar sample should be necessary in order to allow us to calibrate
the relation too as well as the Supernovae Ia (SNeIa) calibration
procedure. Anyway, a first computation of the relation parameters
has been performed by considering the \emph{concordance} model,
namely the Lambda-Cold Dark Matter ($\Lambda$CDM), obtaining a model-dependent luminosity
distance. Unfortunately this procedure leads naturally to the
so-called circularity problem when we take into account a
cosmological use of the GRBs with the Amati relation. A possible
solution has been provided by the use of SNeIa \cite{Perlmutter}.
In other words, one can wonder if it is possible to calibrate GRBs
by adopting at low redshift the SNeIa sample. This proposal has been
already developed in literature by \cite{Liang}. On the other hand,
recently it has been investigated an alternative to solve this
controversy, by adopting a model-\emph{independent} procedure
described by Cosmography, which shall be clarified in the next
section.

\section{The cosmographic Amati relation}

As stressed above, the necessity to account a procedure which is
based on a model-independent way for characterizing the Universe
dynamics is essential; indeed, different cosmological tests may be
taken into account; unfortunately for any case, one of the major
difficulty is related to choosing which may be considered the less
model independent one. One of these, first discussed by Weinberg
\cite{Weinberg} and recently by Visser \cite{Visser}, proposes to
consider the waste amount of kinematical quantities as constraints
to discriminate if a model works well or not. Cosmography is exactly
what we mean for that; we refer to it as the part of cosmology
trying to infer the kinematical quantities as the expansion
velocity, the deceleration parameter and so on, just making the
minimal assumption of a Friedman-Robertson-Walker (FRW) metrics,
being $ds^2=c^2dt^2-a(t)^2\big[dr^2+r^2\left(\sin^2\theta d\theta^2
+ d\phi^2\right)\big]$, \cite{Weinberg}; in particular, it is based
only on keeping the geometry by assuming the Taylor expansion of the
scale factor $a(t)$. In this way we do not do predictions about the
standard Hubble law, but only about its kinematical constraints; it is
worth noting that once expanded as a Taylor series the Hubble law it
is consequent to expand the luminosity distance $d_l$ too and then
the distance modulus $\mu(z)$ \cite{CI2}; unfortunately it is clear
that a similar expansion diverges for $z>1$. Thus to circumvent this
mathematical issue it should be necessary to change the variable,
defining conventionally
\begin{equation}
y = \frac{z}{1+z},
\end{equation}
which limits the redshift range, i.e. $y$ $\in$ (0,1). With this
model-independent formulation we can immediately determine the
cosmographic parameters, in order to reconstruct the trend of the
function $d_l$(y) also at high redshift. Indeed, our aim consists in
assuming the luminosity distance obtained with a good distance
indicators, (SNeIa), extending it also for high redshifts. So far,
as a first step we estimated the cosmographic parameters from a very
large sample of SNeIa, by adopting the Union 2 compilation
\cite{Amanullah}; to perform this, we used a likelihood function
$L$ $\propto$ $e^{-\chi^2 /2}$, where the chi-squared, $\chi^2$ is
given by
\begin{equation}
\chi^2 = \sum_i \frac{(\mu(y) - \mu_i)^2}{\sigma_{\mu_i}^{2}},
\end{equation}
where $\mu_i$ are the distance modulus for each Union SNeIa and
$\sigma\mu_i$ its correspondent error. The results are summarized in
Table \ref{table:no1}.

\begin{figure*}[t!]
\resizebox{\hsize}{!}{\includegraphics[height=6.5cm,width=8.5cm]{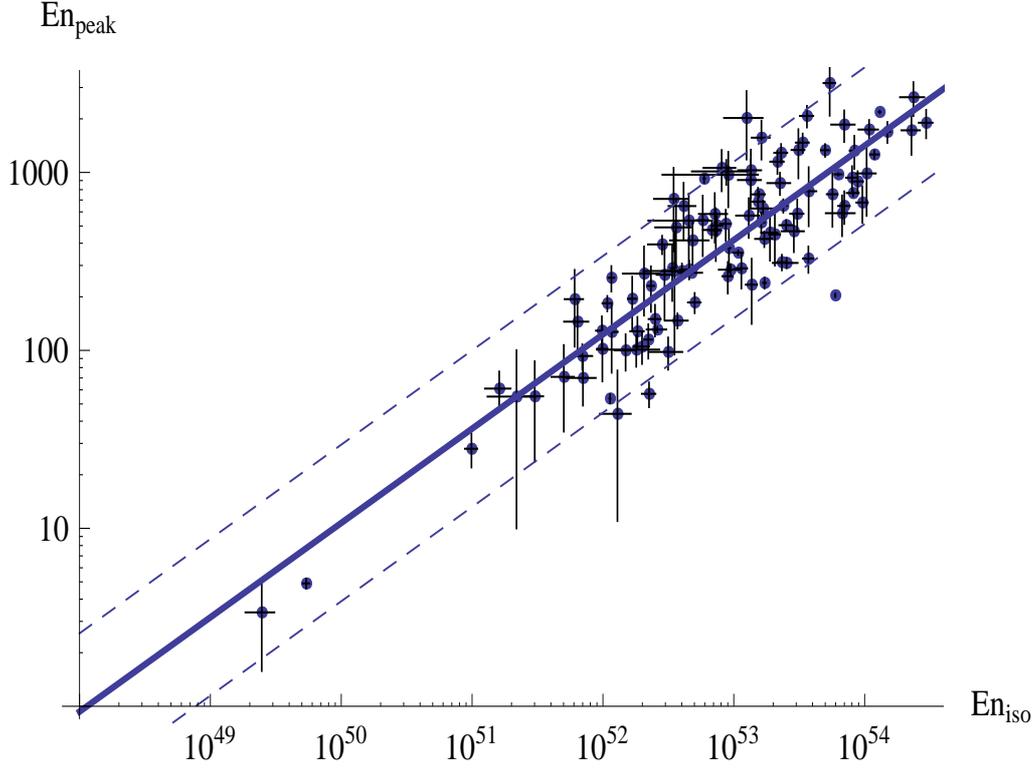}}
\caption{\footnotesize Plot of the cosmographic Amati relation in
the E$_p$ -- E$_{iso}$ one. The line of prediction bounds represents
a deviation of 2$\sigma_{ext}$ from the best fit line, the thick
one.} \label{fig:no1}
\end{figure*}

Once having an expression for $d_l$, in principle, it would be
possible to calibrate the Amati relation too, by using the observed
redshift and the bolometric fluence $S_{bolo}$ of a GRB, computing
the isotropic energy, by inverting eq. \ref{eq:no1}. Then, having as
the Amati relation $E_{iso}$ = $A$ $E_{p,i}^{\gamma}$, we evaluated
the parameters $A$ and $\gamma$, through the use of a sample of 108
GRBs \cite{CI2}, considering as estimator a log-likelihood function
and taking into account the possible existence of an extra
variability $\sigma_{ext}$ of the $y$ data, due to some hidden
variables that we cannot observe directly \cite{D'Agostini}. The
cosmographic calibration gives as results the following values
\begin{equation}
A = 49.17 \pm 0.40, \hspace{1cm} \gamma = 1.46 \pm 0.29,
\end{equation}
and in Fig. \ref{fig:no1} the best fit curve in the E$_p$
-- E$_{iso}$ plane is showed.

\begin{table}
\caption{Cosmographic parameters obtained using the SneIa sample
Union 2. Note that we have considered for the determination of the
jerk $j_0$ and of the snap $s_0$ the flatness condition $\Omega_k$ =
0. The error on $s_0$ does not include the contribute from
covariance terms and the dimension of $H_0$ are $km\,(sMpc)^{-1}$.}
\label{table:no1}
\begin{center}
\begin{tabular}{lcc}
\hline
Parameter & value & error  \\
\hline
H$_0$ & $69.90$ & $0.027$  \\
q$_0$ & $-0.58$ & $0.03$  \\
j$_0$ & $1.50$ & $0.22$  \\
s$_0$ & $-2.96$ & $1.58$  \\
\hline
\end{tabular}
\end{center}
\end{table}

\section{Cosmological applications}

\begin{figure}[]
\resizebox{\hsize}{!}{\includegraphics[width=9cm,
height=6cm]{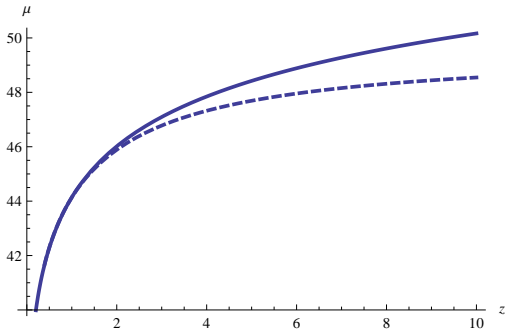}} \caption{ \footnotesize Plot of the
$\mu$(y) computed for a fiducial $\Lambda$CDM cosmological model,
the continuous line, and for the reconstructed $\mu$(y) obtained
from the cosmographic fit of the SneIa, the dashed line, in function
of the z redshift.} \label{fig:no2}
\end{figure}

Although of its elegance, our calibration of the Amati relation has
been obtained using a formulation of $d_l$ which suffers from some
theoretical misleading problems. First of all, since it is defined
for low values of the redshift, the consequent extension to higher
redshift may bring to deviations from the real cosmological picture.
In order to check this discrepancy, we plotted in fig. \ref{fig:no2}
both the distance modulus obtained from Cosmography by SNeIa and
from a flat $\Lambda$CDM paradigm, with $\Omega_{\rho} = 0.27$,
being the matter density. We immediately note a difference of one
magnitude at redshift z $\approx  4$, increasing with $z$, due to
different possible reasons
\begin{description}
  \item[1] the propagation of the systematics in the analysis of the SNeIa used for calibration,
  \item[2] the large scatter in the data sample of the Amati relation,
  \item[3] the standard $\Lambda$CDM model fails at high redshift.
\end{description}
The latter assumption seems to be the less probable one, since the
$\Lambda$CDM model is able to explain the growing of structure
formations too. In addition, in the following, we are going to
present a cosmological application of the GRB sample to estimate the
density parameters.

Let us first compute the isotropic energy E$_{iso}$ for each GRB
from the cosmographic Amati relation, obtaining the distance modulus
for each of them, by using the bolometric fluence S$_{bolo}$ of eq.
(\ref{eq:no2}). Thus, the GRB sample becomes related to the
following theoretical distance modulus
\begin{equation}
D_L(z)= \frac{c}{H_0} {1+z\over
\sqrt{|\Omega_{k}|}}\textrm{sinn}\Big(
\sqrt{|\Omega_{k}|}\int_0^z{d\xi \over E(\xi)} \Big)\,,
\end{equation}
with $E(z)$ the reduced Hubble parameter, i.e.
$E\equiv\frac{H(z)}{H_0}$, while $\Omega_{k}$ represents the
fractional curvature density at $z=0$, and
\begin{displaymath}
{\textrm{sinn} (x)} = \left\{
\begin{array}{ll}
{\textrm{sin}(x)}, & \textrm{if $\Omega_{k}<0$},\\
x, & \textrm{if $\Omega_{k}=0$},\\
{\textrm{sinh}(x)}, & \textrm{if $\Omega_{k}>0$}.
\end{array} \right.
\end{displaymath}

\begin{figure}[]
\resizebox{\hsize}{!}{\includegraphics[width=18cm,height=14cm]{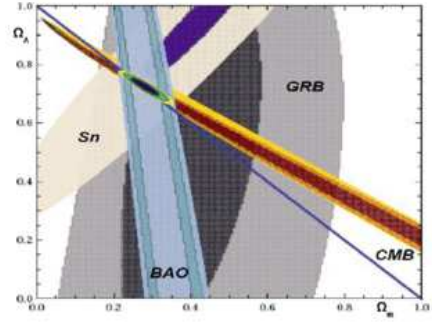}}
\caption{ \footnotesize 68\%, 95\%, and 98\% constraints on
$\Omega_\rho$ and $\Omega_{\Lambda}$ in the $\Lambda$CDM model
obtained from CMB (red), BAO (blue), the Union 2 Compilation (gray
and blue) and the GRB sample considered in this paper (gray and
black). The superimposed contour plot represents the combined final
results.} \label{fig:no3}
\end{figure}

\begin{figure}[]
\resizebox{\hsize}{!}{\includegraphics[width=18cm,
height=14cm]{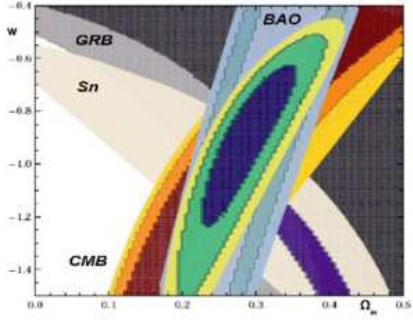}} \caption{ \footnotesize 68\%, 95\%, and 98\%
constraints on $\Omega_\rho$ and $w$ obtained from CMB (orange), BAO
(green), the Union 2 Compilation (gray and blue) and the GRB sample
considered in this paper (gray and black). The superimposed contour
plot represents the combined final results.} \label{fig:no4}
\end{figure}

\begin{figure}[]
\resizebox{\hsize}{!}{\includegraphics[width=9cm,
height=7cm]{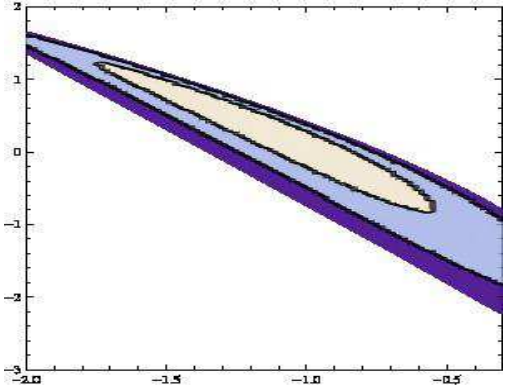}} \caption{ \footnotesize 68\%, 95\%, and 99.7\%
constraints on the CPL parameters $w_0$ and $w_a$ obtained from the
Union 2 Compilation and the GRB sample.} \label{fig:no5}
\end{figure}

We consider as likelihood $L$ $\propto$ $e^{-\chi_{GRB}^2/2}$ the
function given by
\begin{equation}\label{eq:no4}
\chi_{GRB}^2 = \sum_{i=1}^{108} \frac{(\mu_{th} -
\mu_{obs})^2}{\sigma_{\mu,i}^2}\,,
\end{equation}
where $\mu_{obs}$ is the observed distance modulus for each GRB,
with error $\sigma_{\mu,i}$, derived from the Amati relation, while
$\mu_{th}$ is the value of the distance modulus evaluated from the
cosmological model. The constraints have been evaluated by a
combined cosmological test, provided by the SNeIa, baryon acoustic
oscillations (BAO) and cosmic microwave background (CMB). Hence the
total $\chi^2$ is given by \cite{Wang}
\begin{equation}
\chi^2 = \chi^2_{GRB} + \chi^2_{SN} + \chi^2_{BAO} + \chi^2_{CMB}\,.
\end{equation}
In order to perform it, we adopt the Union 2 compilation
\cite{Amanullah}, deriving the constraints and confidence limits by
using the same statistic employed for GRBs. In particular we adopt
the CMB shift parameter $R$
\begin{equation}
R = \sqrt{\Omega_m H_0^2} r(z_{CMB}),
\end{equation}
with $$r(z_{CMB}) = \frac{c}{H_0} |\Omega_k|^{-1/2} \textrm{sinn}
\left(|\Omega_k|^{1/2}\int_0^{z_{CMB}} \frac{d\xi}{E(\xi)}\right)$$
while the $\chi^2$ term is given by
\begin{equation}
\chi^2_{CMB} = \frac{(R - R_{obs})^2}{\sigma_R^2},
\end{equation}
where for R$_{obs}$ and its error we consider the recent WMAP
7-years observations \cite{WMAP}. For the SDSS baryon acoustic
oscillations (BAO) scale measurement and in particular the distance
parameter $A$
\begin{equation}
 A = \Big[r(z_{BAO})^2 \frac{c z_{BAO}}{H(z_{BAO})} \Big]^{1/3} \frac{(\Omega_m H_0^2)^{1/2}}{c
 z_{BAO}}\,,
\end{equation}
with $$r(z_{BAO}) = \frac{c}{H_0} |\Omega_k|^{-1/2} \textrm{sinn}
(|\Omega_k|^{1/2}\int_0^{z_{BAO}} \frac{d\xi}{E(\xi)}),$$ A$_{BAO}$
= 0.469 (n$_s$/0.98)$^{-0.35}$ and $\sigma_A$ = 0.017 \cite{BAO}.
The redshift z$_{BAO}$ = 0.35 while the spectral index is reported
in the respective technical paper \cite{WMAP}. The minimization of
the total $\chi^2$ was done applying a grid-search method in the
parameter space of the model considered. As a first analysis we
considered again the case of the $\Lambda$CDM model, obtaining not
good results, (see fig. \ref{fig:no3}). We conclude that this
happened due to the lacking low-redshift GRB sample, so that we are
not able to give a good accuracy for the best fit values obtained
using the GRB data sample only. A natural extension of $\Lambda$CDM
is represented by the $w$CDM model, the so-called Quintessence model
\cite{Kessence}; here again the results are not in good agreement
with respect what we expected, (see fig. \ref{fig:no4}). In order to
show a good agreement with observations we expect that, since GRBs
are generally at high redshift, a varying Quintessence model, can
provide the trend of the $w$-term, giving rise to a well-fitting
procedure. Among all the possibilities we report below the so-called
Chevallier-Polarski-Linder (CPL) \cite{CPL} as
\begin{equation}
w(z) = w_0 + w_a \frac{z}{1+z},
\end{equation}
where $w\equiv\frac{p}{\rho}$. In this way the distance modulus
curve is sensitive to variations at high redshift of the $w$
quantity, and GRBs are the only source that can shed light on this
topic. The performed analysis developed by using both the SNeIa and
GRB data sample gives results quite in agreement with what we
expected, see fig. \ref{fig:no5} and tab. \ref{table:no2} (together
with the other analysis) and seems to point out GRBs as fundamental
tracers of an evolving Dark Energy equation of state.

\begin{table}
\caption{Final results of our analysis on each cosmological model
considered.} \label{table:no2}
\begin{center}
\begin{tabular}{lcc}
\hline
Parameter & value & error  \\
\hline
$\Lambda$CDM\\
\hline
$\Omega_m$ & $0.279$ & $0.040$  \\
$\Omega_{\Lambda}$ & $0.726$ & $0.034$  \\
$\Omega_k$ & $-0.005$ & $0.001$  \\
\hline wCDM \\\hline
$\Omega_m$ & $0.29$ & $0.08$  \\
$w$ & $-0.87$ & $0.15$  \\
\hline
wCDM + CPL\\
\hline
$\Omega_m$ & $0.27$ & $0.08$  \\
$w_0$ & $-1.18$ & $0.38$  \\
$w_a$ & $0.38$ & $0.59$  \\
\hline
\end{tabular}
\end{center}
\end{table}

\section{Conclusions}

In this work we wondered if the possibility of using GRBs as
distance indicators can be a real resource of the modern
\emph{Precision Cosmology}; obviously this deals with the issue
that, up to now, we cannot claim that GRBs are standard candles. We
developed a statistical (combined) analysis in which the calibration
of the luminosity distance has been performed by a SNeIa sample,
testing different models ($\Lambda$CDM, $w$CDM and CPL
parametrization) with a more complete sample, including GRB data. We
obtain satisfactory results especially in the CPL case. We conclude
that the present data cannot suggest to us something new about the
standard model, but the procedure must be seen as a first
application of the use of GRBs in cosmology, for future
developments. In a next paper we shall present intriguing results,
studying with more accuracy the quoted models and other
alternatives.

\bibliographystyle{aa}

\end{document}